\documentclass[reprint,
superscriptaddress,
nofootinbib,
pre,
]{revtex4-2}

\usepackage{graphicx}% Include figure files
\usepackage{dcolumn}% Align table columns on decimal point
\usepackage{bm}% bold math
\usepackage{hyperref}% add hypertext capabilities
\usepackage{nicefrac}
\usepackage{amssymb}
\usepackage{amsmath}
\usepackage{siunitx}
\usepackage{xcolor}
\newcommand {\LiGS} {$1s^2\,^1\!S_0$ }
\newcommand {\LiTrS} {\ensuremath{2\,^3\!S_1}}
\newcommand {\LiTrP}[1] {\ensuremath{2\,^3\!P_{#1}}}

\def\ketm#1{  \left\vert  #1   \right\rangle   }

\def\sprm#1#2{  \left\langle #1 \left\vert \right. #2 \right\rangle   }

\def\sixjm#1#2#3#4#5#6{  \left\{ \begin{array}{ccc}
                                               #1 & #2 & #3  \\
                                               #4 & #5 & #6
                     \end{array} \right\}   }
\begin{document}

%\preprint{APS/123-QED}

\title{Polarization-Dependent Disappearance of a Resonance Signal:\\ Indication for Optical Pumping in a Storage Ring?}% Force line breaks with \\

\author{W. N\"ortersh\"auser}
 \email{wnoertershaeuser@ikp.tu-darmstadt.de}
 \affiliation{Institut f\"ur Kernphysik, TU Darmstadt, 64289 Darmstadt, Germany%
}%
 \affiliation{Helmholtz Akademie Hessen f\"ur FAIR, GSI Helmholtzzentrum f\"ur Schwerionenforschung GmbH, 64291 Darmstadt, Germany%
}
%  \altaffiliation[Also at ]{Physics Department, XYZ University.}%Lines break automatically or can be forced with \\
\author{A. Surzhykov}%
\affiliation{Physikalisch--Technische Bundesanstalt, 38116 Braunschweig, Germany}
\affiliation{Institut f\"ur Mathematische Physik, Technische Universit\"at Braunschweig, 38106 Braunschweig, Germany}
\affiliation{Laboratory for Emerging Nanometrology Braunschweig, 38106 Braunschweig, Germany}

\author{R. S\'anchez}
 \affiliation{GSI Helmholtzzentrum f\"ur Schwerionenforschung GmbH, 64291 Darmstadt, Germany}
\author{B. Botermann}
  \affiliation{Johannes Gutenberg-Universit\"at Mainz, Institut f\"ur Kernchemie, 55128 Mainz, Germany}
% \author{C. Geppert}
%   \affiliation{Johannes Gutenberg-Universit\"at Mainz, Institut f\"rr Kernchemie, 55128 Mainz, Germany}
\author{G. Gwinner}
  \affiliation{Dept. of Physics \& Astronomy, University of Manitoba, Winnipeg R3T 2N2, Canada}
\author{G. Huber}
  \affiliation{Johannes Gutenberg-Universit\"at Mainz, Institut f\"ur Physik, 55128 Mainz, Germany}
\author{S. Karpuk}
  \affiliation{Johannes Gutenberg-Universit\"at Mainz, Institut f\"ur Physik, 55128 Mainz, Germany}
\author{T. K\"uhl}
 \affiliation{GSI Helmholtzzentrum f\"ur Schwerionenforschung GmbH, 64291 Darmstadt, Germany}
\author{C. Novotny}
  \affiliation{Johannes Gutenberg-Universit\"at Mainz, Institut f\"ur Physik, 55128 Mainz, Germany}
\author{S. Reinhardt}
  \affiliation{Max-Planck-Institut f\"ur Quantenoptik, 85748 Garching, Germany}
\author{G. Saathoff}
  \affiliation{Max-Planck-Institut f\"ur Quantenoptik, 85748 Garching, Germany}
\author{T. St\"ohlker}
 \affiliation{GSI Helmholtzzentrum f\"ur Schwerionenforschung GmbH, 64291 Darmstadt, Germany}
\author{A. Wolf}
  \affiliation{Max-Planck-Institut f\"ur Kernphysik, 69117 Heidelberg, Germany}

\date{\today}% It is always \today, today,

\begin{abstract}
We report on laser spectroscopic measurements on Li$^+$ ions in the experimental storage ring ESR at the GSI Helmholtz Centre for Heavy Ion Research. Driving the  $2s\,^3\!{S}_1\;(F=\nicefrac{3}{2}) \,\leftrightarrow\,2p\,^3\!P_2\;(F=\nicefrac{5}{2}) \leftrightarrow 
2s\,^3\!{S}_1\;(F=\nicefrac{5}{2})$ $\Lambda$-transition in $^7$Li$^+$ with two superimposed laser beams it was found that the use of circularly polarized light leads to a disappearance of the resonance structure in the fluorescence signal. This can be explained by optical pumping into a dark state of polarized ions. We present a detailed theoretical analysis of this process that supports the interpretation of optical pumping and demonstrates that the polarization induced by the laser light must then be at least partially maintained during the round trip of the ions in the storage ring. Such polarized ion beams in storage rings will provide opportunities for new experiments, especially on parity violation.
\end{abstract}

\maketitle

%\tableofcontents

\section{\label{sec:intro}Introduction}

%Short Motivation
Since many years, beams of spin--polarized particles, ions, and atoms have attracted considerable interest both in fundamental and applied science \cite{Kes95, Blu99, Dap18, Mane.2005b}. In atomic, nuclear and particle physics, for example, many experiments have been performed employing electrons, muons, protons and deuterons that can be obtained with high degrees of polarization from thermal to ultrarelativistic energies. 
Less progress has been achieved so far with the production and use of polarized \textit{ion beams}. Here, most of the studies have been restricted to the light--ion and low--energy (keV) domain, where optical pumping is routinely applied \cite{Arnold.1987,Kiefl.2003,Harding.2020}. Only few efforts have been undertaken to produce high--energy beams of polarized lithium and sodium ions at tandem accelerators, e.g. in \cite{Steffens.1981}. Also not much is presently known about beams of spin--polarized ions circulating in storage rings, even though proposals for their production have been discussed already at the end of the last century \cite{Bosch.1997}.  
These beams can be employed for investigating spin--dependent phenomena in atomic and nuclear reactions, measuring the parity violation effects in few--electron systems, and even for the search of a permanent electric dipole moment \cite{Bondarevskaya.2011,Labzowsky.2001,Zolotorev.1997,Ferro.2011}. As an example, let us consider the case of parity non-conservation: In He-like highly charged ions, levels with opposite parity can come very close in energy for a particular $Z$, which enhances parity non-conservation effects. This happens in He-like Eu$^{61+}$, where the $2\, ^3\!P_0$ and the $2\, ^1\!S_0$ states become almost degenerate \cite{Artemyev.2005}. If the system can be prepared in a polarized state, a quenching-type experiment with interference of hyperfine- and weak-quenched transitions can be performed \cite{Labzowsky.2001}. The observable will be the difference in countrate for opposite polarizations. Other examples can be found in \cite{Bondarevskaya.2011}.

Two key issues have to be solved, however, before such experiments with polarized ions in storage rings will become feasible. These are firstly the production of polarized ion beams and secondly the preservation of their polarization during the round trips in the storage ring. The second aspect is a particularly critical issue, as the polarization of ions typically involves the Zeeman substates ($M_F$) of hyperfine levels with large magnetic momenta, given by those of the bound electron. They are, hence, much more sensitive to the fields of the storage device than bare nuclei \cite{Montague.1984} and any excess population in an $M_F$ level is expected to be destroyed rapidly by the fluctuating fields along the stored-ion orbit. 

In the present work we report a joint experimental and theoretical study that addresses both key issues. In particular we consider with a rate equation approach the production and the temporal evolution of hyperfine polarization by Zeeman optical pumping in a two-wavelength excitation scheme. We contrast the theoretical results with a measurement of the steady-state fluorescence intensity from a stored ion beam subjected to this laser excitation, which show a significant dependence on the laser polarization. Within this framework, we conclude that the observations may indicate a preservation of the hyperfine polarization over a considerable number of round trips. To the best of our knowledge, no observations were reported so far that may be interpreted in terms of such extended hyperfine polarization in a storage ring.

The experimental results were obtained within an experiment to test time dilation in Special Relativity with Li$^+$ ions stored in the Experimental Storage Ring (ESR). This experiment probed the fluorescence yield from a three-state, $\Lambda$-type resonance
$2s\,^3\!{S}_1\;(F=\nicefrac{3}{2}) \,\leftrightarrow\,2p\,^3\!P_2\;(F=\nicefrac{5}{2}) \leftrightarrow 2s\,^3\!{S}_1\;(F=\nicefrac{5}{2})$ 
in counterpropagating laser beams along the ion direction. While the $\Lambda$-resonance could indeed be observed and spectroscopically analyzed \cite{Botermann.2014}, we here describe observations showing its depencence on the polarizations on both lasers, which were set as parallel linear polarizations in the previous measurement \cite{Botermann.2014}.
The model calculations are adapted to the parameters of the ESR experiment. 

\section{Lithium-II Level Scheme}

\begin{figure}
    \centering
        \includegraphics[trim = 0 0 0 0, clip,width=\linewidth]{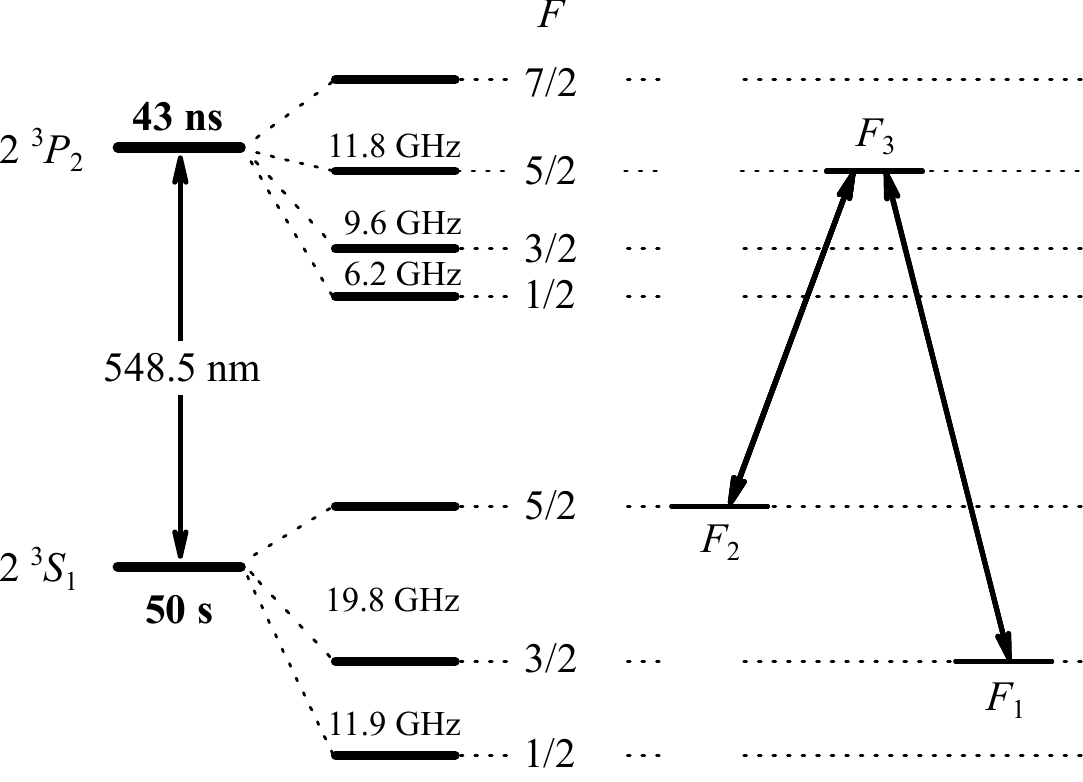}
        \caption{Level scheme and hyperfine structure of the $\LiTrS \rightarrow \LiTrP{2}$ transition in $^7$Li$^+$. The $F_2=\nicefrac{5}{2} \leftrightarrow F_3=\nicefrac{5}{2} \leftrightarrow F_1=\nicefrac{3}{2}$ $\Lambda$ transition  investigated in this work is shown on the right. }
       \label{fig:LiLevelScheme}
\end{figure}

The level scheme of the two-electron system Li$^+$ resembles that of He and is depicted in Fig.\,\ref{fig:LiLevelScheme}. Laser excitation from the \LiGS ground level (not shown) is not possible since the required light is far in the vacuum ultraviolet. But within the triplet system, optical transitions from the metastable $1s2s\,^3\!S_1$ to the $1s2p\,^3\!P_J$ levels ($J=0,1,2$) are accessible at a wavelength of $\lambda_\mathrm{rest}=\SI{548.5}{\nano\meter}$. In the following, we use the abbreviation \LiTrS\ and \LiTrP{J} to denote these levels. The \LiTrS\ level is metastable with a lifetime of approximately 50\,s, sufficiently long for experiments in storage rings. However, the state must be populated already in the ion source. To this end, a Penning ion source (PIG) was used and operated with a CuLi alloy since it was superior to an electron cyclotron resonance ion source (ECRIS) operated with LiF with respect to the amount of metastable ions \cite{Botermann.2011}.
We have used the $\Lambda$-excitation scheme connecting the  $\LiTrS (F=\nicefrac{3}{2},\nicefrac{5}{2})$ and the  \LiTrP{2} $(F=\nicefrac{5}{2})$ hyperfine levels as indicated in the right part of Fig.\,\ref{fig:LiLevelScheme}. For the sake of bookkeeping, below we will enumerate these levels as: $\LiTrS (F_1 = \nicefrac{3}{2})$, $\LiTrS (F_2 = \nicefrac{5}{2})$ and \LiTrP{2} $(F_3=\nicefrac{5}{2})$. 

\section{\label{sec:Setup}Experiment}

\subsection{Ion beam preparation} 

The experimental setup was presented in detail in Refs.\,\cite{Novotny.2009b,Botermann.2011}. 
In brief, $^{7}$Li$^{+}$ ions in the metastable \LiTrS\ level were produced in a PIG source, accelerated in the universal linear accelerator (UNILAC) to an energy of 8~MeV/u and injected into the heavy-ion-synchrotron (SIS) for further acceleration to 58~MeV/u corresponding to a velocity of $\beta=\nicefrac{\upsilon}{c}=0.338$, $(\gamma = 1.0625)$.
Finally, the beam pulses are extracted into the experimental storage ring (ESR) with an efficiency of 10~\%. Typical average ion currents of 20\,$\mu$A or a total number of  $N_\mathrm{ions}=1.3\cdot 10^{8}$ singly charged particles are revolving in the ESR. However, a varying fraction of up to 50\% of the beam was observed to be $^{14}$N$^{2+}$, having a very similar mass-to-charge ratio as $^7$Li$^+$. The residual gas pressure of less than $10^{-11}$\,mbar in the ESR allows for storage times of the ion beam of more than 100\,s. 
The ions are continuously subject to electron cooling to reduce the momentum spread of the ion beam. This reduces the Doppler width of the spectroscopy signal and increases phase-space density and therefore the number of ions that can be addressed simultaneously with a narrowband laser. Approximately 20\,s after injection, the cooling process has reached equilibrium and laser spectroscopy can be performed while electron cooling is continued (see Sec. \ref{subsec:laser_setup}). Additionally, a weak rf signal at the tenth harmonic of the revolution frequency of the ions is applied to an rf-cavity at the ESR to compensate for small energy fluctuations of the electron cooler. This superimposes a bunching potential copropagating with the ions, keeping them precisely at an average velocity defined by the rf frequency and reduces velocity diffusion. 

Laser spectroscopy with full Doppler broadening was performed in the closed $ \LiTrS \; (F=\nicefrac{5}{2}) \rightarrow  \LiTrP{2}\; (F=\nicefrac{7}{2})$ hyperfine transition to locate the resonance \cite{Novotny.2009b,Botermann.2011}. For the cooled beam the Doppler-broadened FWHM was found to be close to 1\,GHz, corresponding to a relative ion momentum spread $\Delta p/p$ close to $1 \cdot 10^{-5}$. It should be noted that a strong ion-laser interaction on such a closed transition can lead to a depletion of a velocity class due to the recoil in repeated photon absorption-emission events. This effect on the external degrees of freedom of an atom or ion is regularly used for laser cooling \cite{Phillips.1982} and has also been employed at storage rings \cite{Schroder.1990,Hangst.1991}. With the laser powers used in our experiment, such effects have not been observed: the closed two-level transition exhibits a regular symmetric Gaussian line shape and for the $\Lambda$-transition the momentum transfer that can occur before the ion is transferred to a dark \LiTrS hyperfine state is very small. Fast cycling will only happen if both lasers talk to the same velocity class and in this case the ion interacts with photons from opposite directions and momentum transfer can again be neglected. Hence, we also do not take into account velocity changes by photon recoil in our theoretical model in Sec.\ \ref{sec:Theory}.

The \LiTrS\ lifetime in the storage ring was determined from the time evolution of the fluorescence signal  to be practically identical with its \mbox{50-s} radiative lifetime and the observed resonance strength allowed us to estimate that the metastable \LiTrS\ level was populated in less than 0.1\% of the Li$^+$ ions.

\subsection{Interaction Region at the ESR}
The ESR is described in detail, e.g., in \cite{Franzke.1987} and with respect to laser spectroscopic experiments in \cite{Nortershauser.2015b}. Here, we will concentrate on the straight interaction region of the ion and the laser beam as it is shown in Fig.\,\ref{fig:ESRSetup}. The ion beam revolving in the ESR is superimposed with the laser beam at the exit of the  60$^\circ$-dipole magnet on the right (ESR's south-west corner) and is separated again at the 60$^\circ$-dipole magnet on the left (ESR's north-west corner). The laser beams are injected through viewports that are mounted in the straight direction at the dipole chambers. Two magnetic quadrupole doublets are located along the straight section, while in the center of the field-free region, a gas target is installed followed by the (fluorescence) detection region used in this experiment. The detection region extends along 65\,cm. The last mirror is about 12-m downstream the ion beam from its first merging point with the laser close to the ESR dipole magnet.

\subsection{Laser setup}
\label{subsec:laser_setup}
\begin{figure}[t]
    \centering
        \includegraphics[trim=0mm 0mm 0mm -10mm, width=8.45cm]{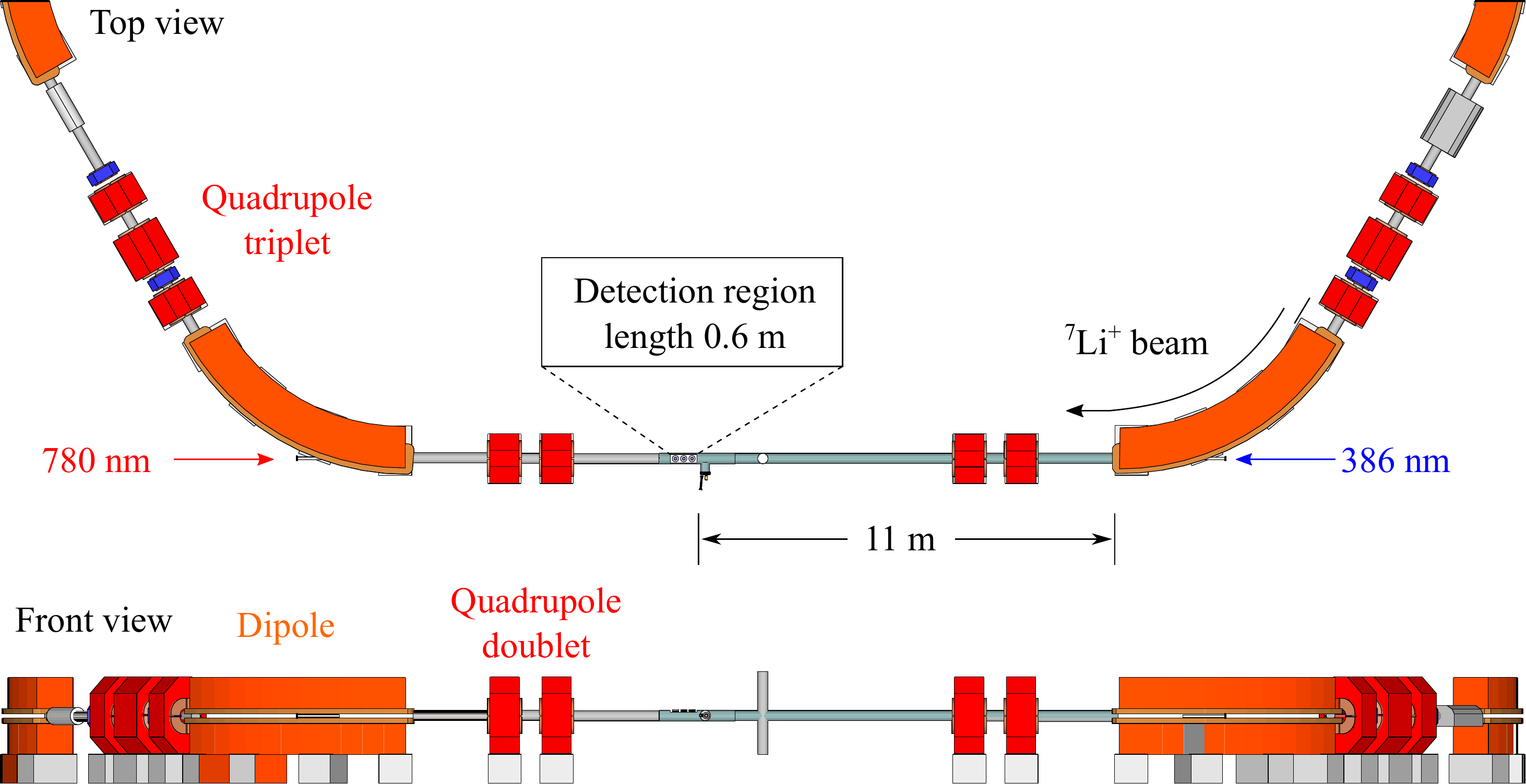}
        \caption{Top-view and front view of the straight section of the ESR where the laser and the ion beam are superimposed. The lasers are directed through the viewports at the 60$^\circ$-dipole magnets to co- and counterpropagate with the ion beam. 
        }
        \label{fig:ESRSetup}
\end{figure}

A Li$^+$ ion in the excited \LiTrP{}\ term decays back to \LiTrS\ with an average lifetime of $\tau(\LiTrP{2})=43$\,ns, which at the beam velocity corresponds to an average fluorescence decay length of $\beta\gamma c \tau=4.52$\,m. To perform Doppler-reduced spectroscopy in the $\Lambda$ scheme described above, two laser beams  were superimposed with the ion beam along the straight section of the ESR, where the fluorescence detection region is located as shown in Fig.\,\ref{fig:ESRSetup}. 
As the average fluorescence length is to short for a separation between ion beam and laser beam before detection, the lasers must traverse the detection region and background from scattered laser light cannot be avoided \cite{Nortershauser.2015b}. The Doppler effect at 33.8\,\% of the speed of light shifts the excitation wavelength from $\lambda_{0}=548.5$\,nm at rest to $\lambda_{p}=386$\,nm for the laser beam copropagating (parallel) with the ions, while excitation with the counterpropagating (antiparallel) laser beam has to be performed at a wavelength of $\lambda_{a}=780$\,nm in the laboratory frame. 

% Laser system
Copropagating laser light is produced with a commercial tunable solid-state titanium-sapphire (Ti:Sa) ring laser (coherent, 899-21) operated at 772\,nm pumped by a Nd:YVO$_4$ laser (Coherent, Verdi) at 532\,nm. Frequency doubling in an enhancement cavity (Wavetrain2, Sirah) with a BBO ($\beta$-BaB$_2$O$_4$) crystal provided the required wavelength at \SI{386}{\nano\meter}. The Ti:Sa laser is stabilized to the rovibronic P(42)1-14 transition in molecular iodine $^{127}$I$_2$ by frequency modulation saturation spectroscopy \cite{Hall.1981,Reinhardt.2007}. The linewidth of the laser is approximately 250\,kHz on a \mbox{1-s} integration time. It is estimated that this increases to about 350\,kHz after frequency doubling.
The counterpropagating laser beam is generated with a pair of diode  lasers in Littrow configuration (Toptica, DL100) of which one is locked to a hyperfine transition at 780\,nm in the D2 line of $^{87}$Rb \cite{Barwood.1991}, while the second one is tunable and referenced to the first one by a frequency-offset lock \cite{Schunemann.1999}. The linewidth of the counterpropagating laser is about 2\,MHz. More details are provided in \cite{Novotny.2009b,Botermann.2014}.

%Laser Transport 
Both lasers are transported to the storage ring using polarization-maintaining optical fibers. For the UV light for collinear excitation a 15-m long  Thorlabs fiber (type PM-S350-HP) is used, while the IR light requires a fiber with a length of 50\,m (type PM780-HP). 

%Coupling to beamline
The fused-silica viewports at the ESR's dipole magnets have a surface flatness  of $\lambda$/20 and can be tilted to send the reflections of the window surfaces out of the ESR to reduce stray light. The angles and positions of the laser beams are controlled with accuracy of $\Delta \vartheta = \SI{20}{\micro\radian}$ and $\Delta x \approx \SI{50}{\micro\meter}$ by motorized rotation and translation stages. Two scraper pairs, situated in the middle of the experimental section, 6.5 meters apart from each other are used to establish ion beam -- laser beam overlap inside the vacuum chamber. The scrapers can be introduced into the beam path with a reproducibility of better than 0.1\,mm, allowing us to guarantee parallel beams within \SI{80}{\micro\radian}. 

Laser beam parameters have been determined with a beam viewer: both foci are located 17 m downstream of the detection regions last mirror, actually outside the ESR beamline, with radii (at $\nicefrac{1}{e^2}$ intensity) of 1.10(5)\,mm and 2.71(10)\,mm for the blue and red laser beams, respectively. Along the \mbox{11-m} long interaction path prior to detection, the beam radii vary accordingly from 3.8\,mm (3.4\,mm) at the dipole magnet down to 3.1\,mm (2.2\,mm) at the detection region for the red (blue) laser. The laser power was adjusted to 5\,mW for the red and 0.4\,mW for the blue laser. This corresponds to intensities of roughly 3.3 to 4.9 times the saturation intensity of ${\mathcal I}_0=6.7\,\mathrm{mW/cm}^2$ \cite{Wanner.1998} for the red and 30-80\% saturation intensity for the blue transition along the interaction path.  In the theoretical simulation (see Sec.\,\ref{sec:Theory}), we use an average size of 3.5\,mm (red) and 2.8\,mm (blue), respectively, and transform the intensity into the rest frame of the ion. These values are listed in Tab.\,\ref{tab1}.

\subsection{Fluorescence Detection}
The system for fluorescence detection is situated in the middle
of the experimental section. It was originally built for laser spectroscopy on hydrogen-like Pb \cite{Seelig.1998}, a detailed description is provided in \cite{Sanchez.2017} and its detection properties were modelled in \cite{Hannen.2013}.
Three photomultiplier tubes (PMTs) (two \textit{Hamamatsu}, R2256P and one \textit{Thorn EMI, 9635QB}) are used to detect the fluorescence photons. All PMTs are equipped with cooler housings to keep the temperature below $5\,^{\circ}\mathrm{C}$ to reduce thermal noise. Each PMT is additionally equipped with a combination of lenses optimized to focus photons from the beam position onto the photocathode. The detection system can only collect light that is emitted under angles of at least 20$^\circ$ with respect to the beam direction. The fluorescence detection is therefore restricted to a wavelength range between 410\,nm and about 700\,nm with a combination of optical filters (Schott, BG~39 \& Itos KV408) to suppress laser stray light \cite{Botermann.2011}. Consequently, processes that would steer the emission direction of the ions away from these forward directions could lead to a signal reduction even with a constant fluorescence rate.

\subsection{Spectroscopic Procedure}

Resonance spectra of the Li$^+$ ions are recorded in the following way: First, the Ti:Sa laser producing the copropagating beam (wavelength $\lambda_p$) is stabilized to the iodine line and the ion beam energy is adapted by changing the electron energy at the electron cooler to resonantly excite the ions along the  $\LiTrS (F_1 = \nicefrac{3}{2}) \rightarrow \LiTrP{2} (F_3 = \nicefrac{5}{2})$ transition. Then, the $\Lambda$-resonance signal is observed by tuning the frequency of the counterpropagating (diode) laser beam (wavelength $\lambda_a$) across the  $\LiTrS (F_2 = \nicefrac{5}{2}) \rightarrow \LiTrP{2} (F_3 = \nicefrac{5}{2})$ transition. 
While $\lambda_{a}$ is scanned, the fluorescence intensity is recorded for four phases when the laser beams are switched on or off ($\lambda_a$ only, $\lambda_p$ only, both lasers, and no laser) \cite{Botermann.2014} and the $\Lambda$-resonance spectrum is obtained from the difference between the ``both-lasers'' phase and the summed signals of the two single-laser phases.
In the fluorescence signal of the ``$\lambda_a$-on'' phase, a Doppler-broadened spectrum occurs through $F_2=\nicefrac{5}{2} \rightarrow F_3=\nicefrac{5}{2}$ excitation for a velocity-class of the ions selected by the narrow-band laser. This fluorescence signal is reduced by the optical pumping to the $F_1=\nicefrac{3}{2}$ hyperfine levels through spontaneous emission $F_3 \rightarrow F_1$.
In the ``both-lasers'' phase and in the signal difference representing the $\Lambda$
resonance spectrum, an additional spectral structure occurs when $\lambda_a$ selects for the $F_2 \rightarrow F_3$ transition the same velocity class as the fixed $\lambda_p$ laser selects for the $F_1\rightarrow F_3$ transition. Hyperfine pumping to the $F_1$ level is then avoided so that the average population in $F_3$ and, correspondingly, the fluorescence intensity rises.
Indeed, a $\Lambda$-resonance signal at a width significantly reduced compared to the Doppler width could be observed and spectroscopically analyzed to yield the desired test of Special Relativity \cite{Botermann.2014}.
The remaining linewidth of about 100\,MHz is still more than an order of magnitude larger than the natural linewidth of approximately 3.7\,MHz. This is ascribed to velocity-changing collisions happening during many revolutions in the ESR, shifting ions dark-pumped by one of the lasers back into resonance with the other one \cite{Botermann.2014}.

\begin{figure}[t]
    \centering
        \includegraphics[trim=0mm 0mm 0mm -10mm, width=8.45cm]{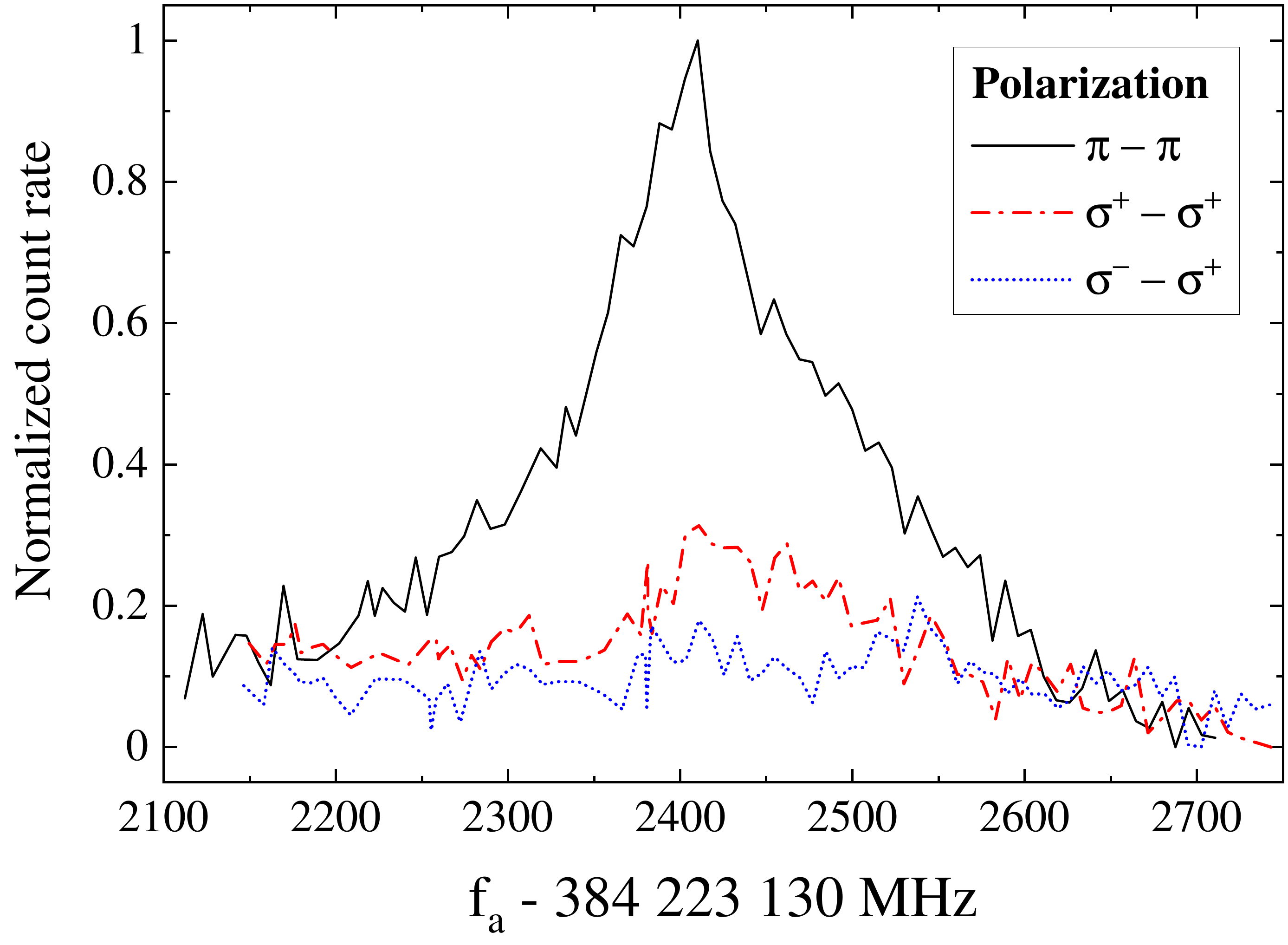}
        \caption{Optical resonances of Li$^+$ ions in the ESR recorded with different laser polarizations during the ``both lasers on'' phase (see text).
        The linear-linear case was recorded with both laser beams polarized in the horizontal (ring) plane.
        For  circularly polarized light, the helicities of the UV ($1^\mathrm{st}$) and IR photons ($2^\mathrm{nd}$) are indicated. The $x$-axis represents the frequency of the IR light ($\lambda_a$) produced by the diode laser that counterpropagates to the ion beam and drives the $\LiTrS (F_2 = \nicefrac{5}{2}) \rightarrow \LiTrP{2} (F_3 = \nicefrac{5}{2})$ transition. The UV light ($\lambda_p$) of the Ti-sapphire lasers second harmonic is copropagating and its frequency is fixed on the $\LiTrS (F_1 = \nicefrac{3}{2}) \rightarrow \LiTrP{2} (F_3 = \nicefrac{5}{2})$ resonance of the central velocity group of the revolving ions.}
        \label{fig:Resonances}
\end{figure}

\section{\label{sec:Results}Results}
In the experimental runs leading to the Relativity Test \cite{Botermann.2014}, also the effect of the laser polarizations on the $\Lambda$-signal was studied. In particular, the situation of parallel linear polarizations, chosen for the measurement, was replaced by chosing circular polarizations for both lasers. With this goal, $\nicefrac{\lambda}{4}$-plates in front of the ESR windows were introduced and their fast axis rotated to obtain circular polarization. Here and in the following, we always refer to the helicity state of the photon, i.e.\ the projection of the photon spin onto its own momentum direction.

The signals (denoted as $\Lambda$-spectra in the following) observed when both lasers are on are shown in Fig.\,\ref{fig:Resonances} for the tested laser polarizations.
The black solid curve represents the resonance with both lasers linearly polarized in the horizontal plane\footnote{A similar  signal for vertical polarization was also recorded.} ($\pi$--$\pi$). This signal shows a sharp structure representing the suppression of hyperfine-population pumping by a single laser, as described above and analyzed in Ref.\,\cite{Botermann.2014}.
A wider (Doppler-broadened) pedestal of this resonance occurs in addition. It was explained by the diffusion of ion velocities during their round trips, which leads to a replenishment of ions in the laser-accessible state from neighboring velocity classes with state populations that have not been reduced by the scanned (IR) laser. Similarly, hyperfine population pumping by the $\Lambda$-type interaction with both lasers can be suppressed over a wider range of $\lambda_a$ detunings which explains the increased linewidth of the observed $\Lambda$-signal. Such velocity changes can occur in electron-ion collisions in the electron cooler, in ion-atom collisions with residual gas atoms or can be induced by the rf cavity used for the weak bunching.
The dash-dotted trace in red was recorded with both laser beams having $\sigma^+$ polarization. Here, the fluorescence near the $\Lambda$-resonance becomes much smaller, regarding both the pedestal and the resonant structure. Only a weak signal above background remains.
Then, the $\nicefrac{\lambda}{4}$-plate of the UV laser was rotated by $90^\circ$ to obtain $\sigma^-$ light. In this case, even the weak resonance vanishes, as displayed by the blue dotted line.

\begin{figure}[t]
    \centering
        \includegraphics[trim=0mm 0mm 0mm 0mm, width=8.45cm]{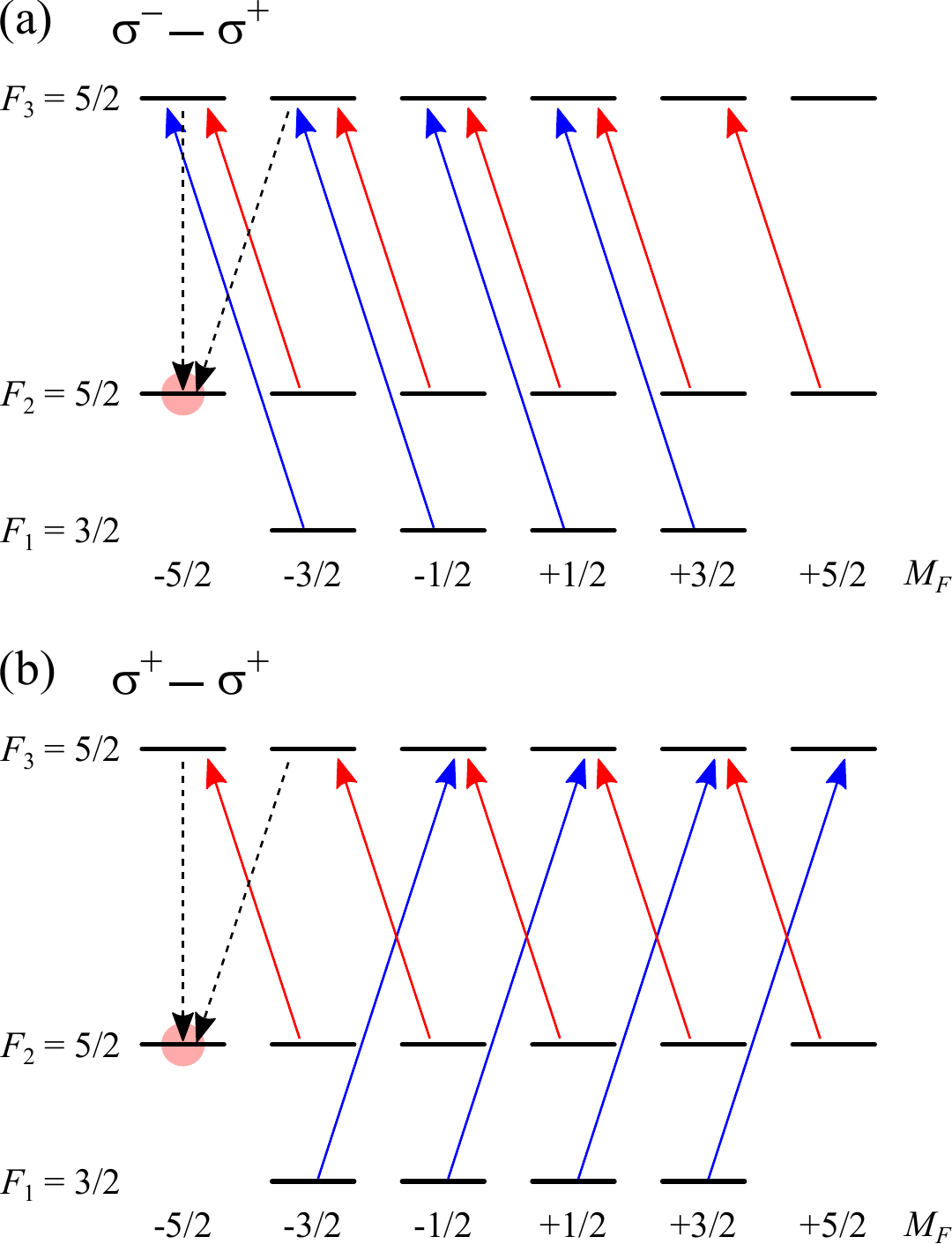}
        \caption{Level scheme with $M_F$ substates for the explanation of the dark state in the $\Lambda$-scheme with two circularly polarized laser beams. The upper panel (a) displays the case of $\sigma^-$(UV)--$\sigma^+$(IR) polarization, while the lower panel (b) corresponds to the $\sigma^+$(UV)--$\sigma^+$(IR) case.}
        \label{fig:DarkStates}
\end{figure}

We interpret the disappearence of the fluorescence near the $\Lambda$-resonances even under the combined effect of both lasers $\lambda_a$ and $\lambda_p$ by optical pumping between $M_F$ Zeeman sublevels with respect to the laser and ion-beam direction as a quantization axis, as illustrated in Fig.\,\ref{fig:DarkStates}. For the $\sigma^--\sigma^+$ case (Fig.\,\ref{fig:DarkStates} (a)), $\Delta M_F=+1$ transitions are indicated by the solid lines in blue and red, respectively. From all $M_F$ states of the two lower levels of the $\Lambda$-transition, only the  %$\ketm{\LiTrS:F_2 = \nicefrac{5}{2},\,M_F=-\nicefrac{5}{2}}$
$M_F=-\nicefrac{5}{2}$ state of the $\LiTrS(F_2 = \nicefrac{5}{2})$ level cannot be excited with $\sigma^-$ IR light, but this state is populated by partial decays from the $M_{F} = -\nicefrac{3}{2},-\nicefrac{5}{2}$ substates of the $\LiTrP{2}(F_3 = \nicefrac{5}{2})$ level as indicated by the dashed arrows.
All other decays leading back to laser-coupled $M_F$ states are not shown. Both lasers are shifting the population towards this ``dark'' substate. In absence of a redistribution process between the different $M_F$ states of a level, resonant laser excitation must eventually end with all population accumulated in the dark substate and fluorescence will stop. Similar qualitative analysis, that takes into account transitions between various hyperfine substates $M_F$, may explain the existence of a weak fluorescence signal, observed for the $\sigma^+-\sigma^+$ case. As seen from Fig.\,\ref{fig:DarkStates} (b), in this case the UV laser excites $\Delta M_F=+1$ transitions. Again, the $M_{F} = -\nicefrac{5}{2}$ magnetic substate of the $\LiTrS(F_2= \nicefrac{5}{2})$ level becomes the dark substate but this time the excitation with the UV beam shifts the population in the opposite direction, $\Delta M_F = +1$. Hence, one can expect the process to slow down and that it will take longer until fluorescence vanishes.

It should be noted that the case denoted as a $\pi$--$\pi$ polarization of both lasers, for the $M_F$ states referring to the lasers' direction represents a coherent superposition of the transitions marked in both panels in Fig.\,\ref{fig:DarkStates}.
Hence, repumping from the dark states is ensured and a large $\Lambda$ repumping effect is observed.

In both cases when the lasers are circularly polarized, the population of the $M_F=-\nicefrac{5}{2}$ dark substate leads to an effective pumping of the hyperfine-level population to $F_2$ while state $F_1$ is depleted. However, this optical pumping is also connected with a steady-state \textit{polarization} of the stored $F_2$ ions in the $M_F=-\nicefrac{5}{2}$ state.
This consequence of the observed laser-polarization dependence was not followed further in the Special Relativity test experiment. However, as pointed out in Sec.\,\ref{sec:intro}, it could have exciting consequences such that we consider this effect more closely in the following.
%With this explanation, further trials to investigate this issue were abandoned during the beamtime and the test of Special Relativity was carried out using linearly polarized light as reported in \cite{Botermann.2014}. Only later, the implication of this disappearance was recognized: If this happened indeed, a polarized ion beam was produced. Such a possibility would be exciting as discussed in the introduction. However, there are two scenarios, how this can happen:

For the described polarization effect to yield experimental consequences, two scenarios can be considered:
\begin{itemize}
    \item [(\textit{i})] The laser beam interacts  with the ions along a distance of 11\,m in front of the detection region, which corresponds to $\sim 2.4$ spontaneous decay lengths. Hence, the polarization might build up already along this distance. Even a partial polarization build-up may sufficiently change the emission characteristic of the ions so that it is steered into directions where the mirror system is less sensitive, i.e., outside of the angular range between $30^\circ$ and $80^\circ$ with the highest efficiency for detection. 
    We denote this as the ``fast pumping'' scenario, in which any polarization in the ion beam is considered to be again destroyed during a round trip. Conservation of the polarization during the ion storage is irrelevant in this scheme. 
    \item [(\textit{ii})] In another
scenario the buildup of polarization according to scheme (\textit{i}) would not be sufficient to explain the observations. Then, at least a part of the polarization must be conserved after the round trip such that the polarization can slowly build up over several round trips. 
\end{itemize}
In the following theoretical section, we will investigate which of these two scenarios is more suitable for explaining our observations. Therefore, we analyze the laser-ion interaction with a rate-equation approach, taking all relevant experimental boundary conditions into account.

\section{\label{sec:Theory}Theory}

\subsection{System of rate equations for Li$^{+}$ ions}
\label{subsec:rate_equations}

In order to better understand the experimental results from the previous section, we have performed a theoretical analysis of the population dynamics of Li$^+$ ions in the presence of two superimposed laser fields. Generally, such an analysis would require solution of the Liouville--von Neumann equation that describes the time evolution of the ionic density matrix \cite{Auzinsh.2010}. For 
a discussion as it is intended here, just aiming at a decision which pumping scheme introduced above is supported by theory, 
we can restrict ourselves to a less computationally demanding approach based on a system of rate equations. The rate--equation approach has been widely employed in the past for the investigation of the optical pumping of ions in storage rings \cite{Prozorov.2003, Bondarevskaya.2011}. It predicts how the populations of magnetic substates of an ion, as described by the \textit{diagonal} elements of the density matrix, evolve over the time due to transitions from/to another substates. The focus on the substate populations implies neglecting the coherence between the ionic (sub--) states, which is characterized by the non--diagonal density--matrix elements. This assumption is well justified for the analysis of the present ESR experiment in which fast spontaneous decay of the excited state $\LiTrP{2}$ as well as ionic collisions and interactions with the external fields in the ring lead to the loss of the coherence.

In the present work we solve the system of coupled differential equations:
\begin{widetext}
\begin{subequations}\label{eq:rate_equations_general}
\begin{eqnarray}
    \frac{{\rm d}N_{F_1, M_{F_1}}}{{\rm d}t} &=& 
    - N_{F_1, M_{F_1}} \sum\limits_{M_{F_3}} \Gamma^{{\rm \, exc}}_{F_1 M_{F_1}, F_3 M_{F_3}} + \sum\limits_{M_{F_3}} N_{F_3, M_{F_3}}  \Gamma^{{\rm \, dec}}_{F_3 M_{F_3}, F_1 M_{F_1}} \, , \\[0.2cm]
    \frac{{\rm d}N_{F_2, M_{F_2}}}{{\rm d}t} &=& 
    - N_{F_2, M_{F_2}} \sum\limits_{M_{F_3}} \Gamma^{{\rm \, exc}}_{F_2 M_{F_2}, F_3 M_{F_3}} + \sum\limits_{M_{F_3}} N_{F_3, M_{F_3}}  \Gamma^{{\rm \, dec}}_{F_3 M_{F_3}, F_2 M_{F_2}} \, , \\[0.2cm]    
    \frac{{\rm d}N_{F_3, M_{F_3}}}{{\rm d}t} &=& 
    - N_{F_3, M_{F_3}} \left( \sum\limits_{M_{F_1}} \Gamma^{{\rm dec}}_{F_3 M_{F_3}, F_1 M_{F_1}} + \sum\limits_{M_{F_2}} \Gamma^{{\rm dec}}_{F_3 M_{F_3}, F_2 M_{F_2}} \right) \nonumber \\ 
    &+& 
    \sum\limits_{M_{F_1}} N_{F_1, M_{F_1}} \Gamma^{{\rm exc}}_{F_1 M_{F_1}, F_3 M_{F_3}}\, +
    \sum\limits_{M_{F_2}} N_{F_2, M_{F_2}} \Gamma^{{\rm exc}}_{F_2 M_{F_2}, F_3 M_{F_3}} , 
\end{eqnarray}
\end{subequations}
\end{widetext}
that describes the populations $N_{F, M_F}(t)$ of the hyperfine substates $\ketm{F_1,M_{F_1}} \equiv \ketm{\LiTrS: F_1 = \nicefrac{3}{2}, \, M_{F_1}}$, $\ketm{F_2, M_{F_2}} \equiv \ketm{\LiTrS: F_2 = \nicefrac{5}{2}, \, M_{F_2}}$ and $\ketm{F_3, M_{F_3}} \equiv \ketm{\LiTrP{2}: F_3 = \nicefrac{5}{2}, \, M_{F_3}}$ of Li$^+$ ions. Here $F$ is the total angular momentum of the ion and $M_F$ is its projection onto the beam propagation direction, chosen as the quantization axis. 

In Eqs.~(\ref{eq:rate_equations_general}), moreover,
$\Gamma^{{\rm \, exc}}_{F_i M_{F_i}, F_3 M_{F_3}}$ is the rate for the photoexcitation $\ketm{F_i, M_{F_i}} + \hbar\omega \to \ketm{F_3, M_{F_3}}$, while $\Gamma^{{\rm dec}}_{F_3 M_{F_3}, F_i M_{F_i}} = \Gamma^{{\rm dec, sp}}_{F_3 M_{F_3}, F_i M_{F_i}} + \Gamma^{{\rm dec, ind}}_{F_3 M_{F_3}, F_i M_{F_i}}$
with $i=1,2$ is the sum of the spontaneous and laser--induced decay rates. These rates for transitions between particular \textit{hyperfine} substates can be easily traced back to the rate of the spontaneous $2\, ^3\!P_2 \to 2\, ^3\!S_1$ \textit{fine--structure} decay that we denote as $\Gamma_{2\, ^3\!P_2, \, 2\, ^3\!S_1}$. For example, with the help of the angular momentum algebra one can find
\begin{eqnarray}
    \label{eq:spontaneous_decay_rate}
    \Gamma^{{\rm dec, sp}}_{F_3 M_{F_3}, F_i M_{F_i}} &=& 5 (2F_i + 1) \, 
    \sprm{F_i M_{F_i} \, L M}{F_3 M_{F_3}}^2 \nonumber \\
    &\times& \sixjm{F_3}{F_i}{L}{1}{2}{I}^2 \, \Gamma_{2\, ^3\!P_2, \, 2\,^3\!S_1} \, ,
\end{eqnarray}
where $I = 3/2$ is the nuclear spin, and the electric--dipole approximation, $L = 1$, for the photon--ion interaction is assumed. By employing this expression and the theory of Einstein coefficients, we obtain next the rate for the laser--induced decay as
\begin{eqnarray}
    \label{eq:induced_decay_rate}
    \Gamma^{{\rm dec, ind}}_{F_3 M_{F_3}, F_i M_{F_i}} &=& 5 (2F_i + 1) \, \sixjm{F_3}{F_i}{L}{1}{2}{I}^2 \nonumber \\
    &\times& \left(\sum_{p = \pm 1}  c_{p} \, p \, \sprm{F_i M_{F_i} \, L\, p}{F_3 M_{F_3}} \right)^2 \nonumber \\
    &\times& \left( {\mathcal I} \frac{\pi^2 c^2}{\hbar \omega^3 \Delta\omega} \right) \, \Gamma_{2\, ^3\!P_2, \, 2\,^3\!S_1} \, .
\end{eqnarray}
Here, ${\mathcal I}$, $\omega$ and $\Delta\omega$ are the intensity, angular frequency and (frequency) width of the incident laser radiation, whose values are displayed in Table~\ref{tab1}. Moreover, the coefficients $c_p$ characterize the photon polarization vector ${\bm \epsilon}$, written in the helicity ($p = \pm 1$) representation as
\begin{equation}
    {\bm \epsilon} = c_{1} {\bm \epsilon}_{1} + c_{-1} {\bm \epsilon}_{-1} \, ,
    \: \: \: \left| c_1 \right|^2 + \left| c_{-1} \right|^2 = 1 \, . 
\end{equation}
By choosing these parameters as $c_{1} = c_{-1} = 1/\sqrt{2}$ and $c_{\pm 1} = 1$ one can ``construct'' $\pi$--linearly-- and $\sigma^{\pm}$--circularly polarized light, respectively. Finally, the photoexcitation rate $\Gamma^{{\rm \, exc}}_{F_i M_{F_i}, F_3 M_{F_3}}$ is directly related to $\Gamma^{{\rm dec, ind}}_{F_3 M_{F_3}, F_i M_{F_i}}$ by the principle of the detailed balance.

\begin{table}
\begin{ruledtabular}
\begin{tabular}{ll}
Intensity of the blue laser  & ${\mathcal I}_{\rm blue} = 1.6$~mW/cm$^2$ \\[0.2cm]
Intensity of the red laser  & ${\mathcal I}_{\rm red} = 52.5$~mW/cm$^2$ \\[0.2cm]
Laser angular frequency & $\omega$~=~3.4~$\times 10^{15}$~Hz \\[0.2cm]
Laser width of the blue laser & $\Delta\omega_{\rm blue}$~=~1.5~$\times 10^{6}$~Hz \\[0.2cm]
Laser width of the red laser & $\Delta\omega_{\rm red}$~=~1.8~$\times 10^{7}$~Hz \\[0.2cm]
Rate of the $2\, ^3\!P_2 \to 2\, ^3\!S_1$ decay & $\Gamma_{2\, ^3\!P_2, \, 2\,^3\!S_1} = 2.2 \times 10^7$~Hz 
\end{tabular}
\end{ruledtabular}
\caption{\label{tab1} Parameters of the incident laser radiation and of the Li$^+$ ion used in the evaluation of the transition rates. All parameters are given in the rest frame of an ion, moving with the relativistic velocity $\beta$~=~0.338.}
\end{table}
\begin{figure}[t!]
        \includegraphics[width=8.5cm]{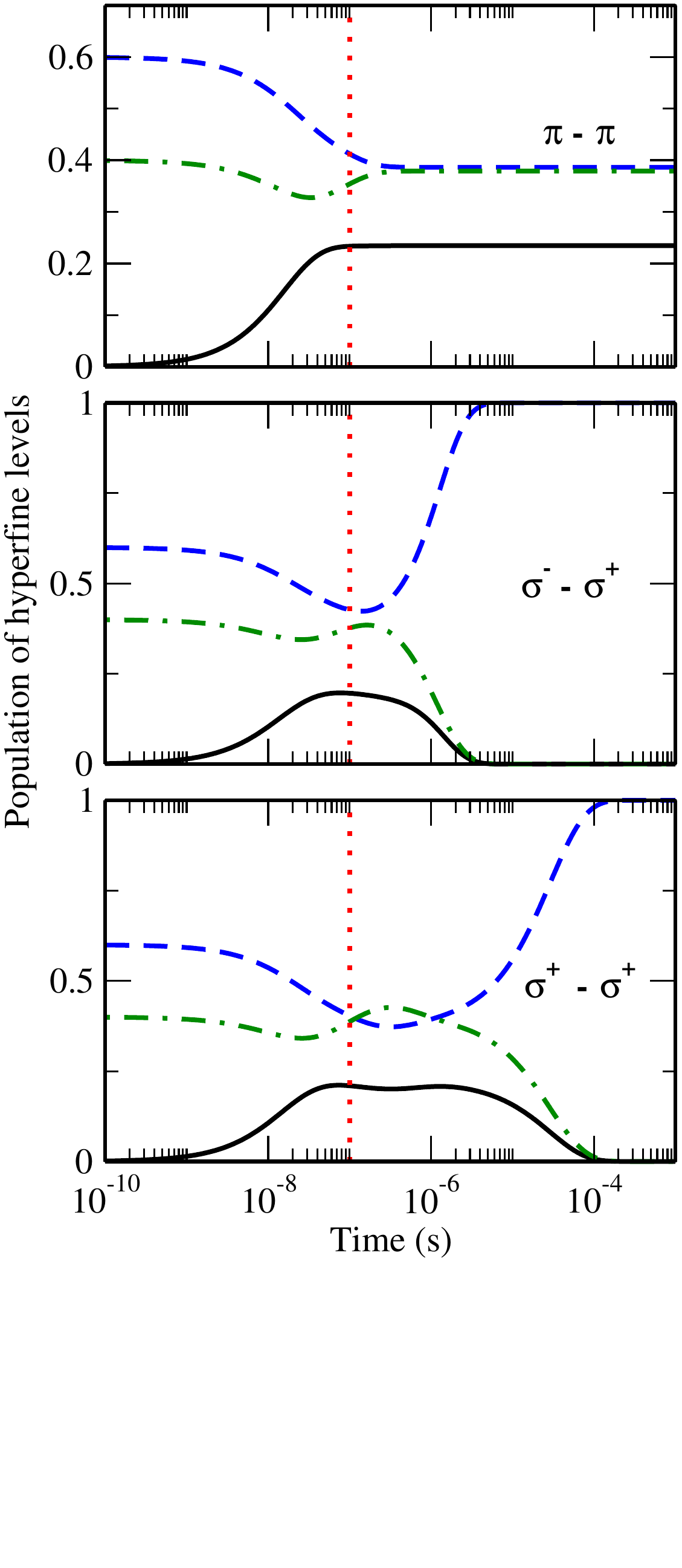}
        \vspace*{-4.0cm}
        \caption{The relative population of the $\LiTrS: F_1 = \nicefrac{3}{2}$ (green dash--dotted line), $\LiTrS: F_2 = \nicefrac{5}{2}$ (blue dashed line) and $\LiTrP{2}: F_3 = \nicefrac{5}{2}$ (black solid line) hyperfine levels of Li$^+$ ions as a function of time. The calculations have been performed for two counter--propagating laser beams that are both either linearly polarized in the horizontal plane (upper panel) or circularly polarized. For the latter case we consider two beams with the opposite $\sigma^{-}-\sigma^{+}$ (middle panel) and the same $\sigma^{+}-\sigma^{+}$ (lower panel) polarizations. Finally, the red dotted vertical line marks the time $t_{\rm det}$, that Li$^{+}$ ions, moving with velocity $v = 0.338 \, c$, need to reach the center of the detection region from the point where they enter the laser--overlap region.}
        \label{Fig1_theo}
\end{figure}

\begin{figure*}[t]
        \includegraphics[width=17.5cm]{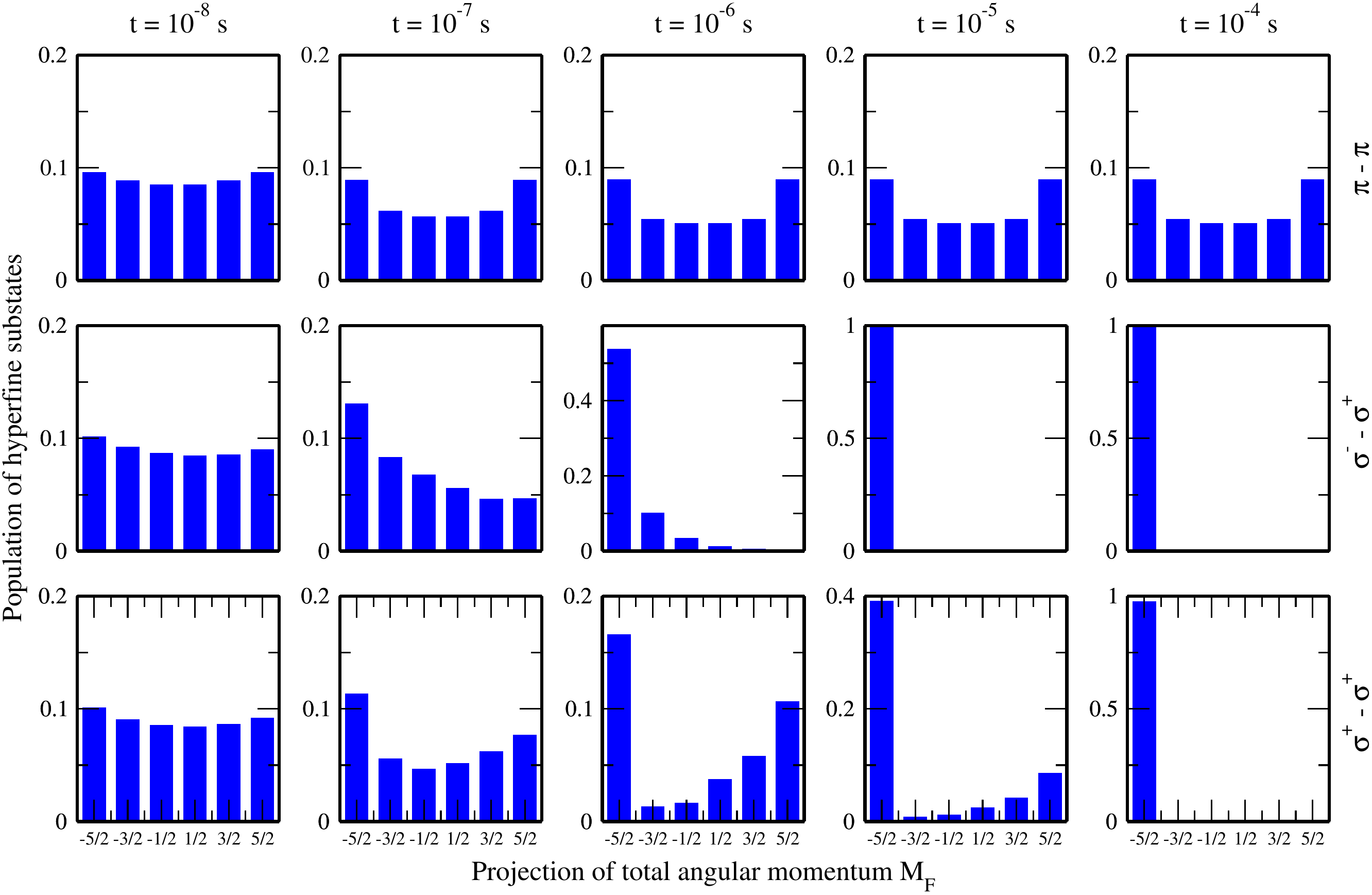}
        \vspace*{0.0cm}
        \caption{The relative population $N_{F_2 M_{F_2}}$ of the hyperfine substates $\ketm{\LiTrS: F_2 = \nicefrac{5}{2}, \, M_{F_2}}$ at times $t = 10^{-8}$~s, $t = 10^{-7}$~s, $t = 10^{-6}$~s, $t = 10^{-5}$~s and $t = 10^{-4}$~s after initially unpolarized Li$^+$ ions enter the laser--overlap region. Calculations have been performed for two counter--propagating laser beams that are both either linearly polarized in the horizontal plane (upper panel) or circularly $\sigma^{-}-\sigma^{+}$ (middle panel) and $\sigma^{+}-\sigma^{+}$ (lower panel) polarized. For the latter two polarization cases one can observe the formation of the dark substate $\ketm{\LiTrS: F_2 = \nicefrac{5}{2}, \, M_{F_2}=-\nicefrac{5}{2}}$. This formation, however, takes much longer time for the $\sigma^{+}-\sigma^{+}$ polarization case where blue and red lasers transfer the substate populations in opposite directions, see lower panel of Fig.~\ref{fig:DarkStates}.}
        \label{Fig_3S1_sublevels}
\end{figure*}

In what follows we will apply the developed theoretical approach to investigate the population dynamics of Li$^+$ hyperfine levels for two scenarios. Thus, in Sec.~\ref{subsec:single_cycle} the naive \textit{single--run} scenario will be considered in which the ions enter the laser--overlap region at $t = 0$ and move in it infinitely long. This academic case will allow us to determine the characteristic time--scales of the population dynamics and to discuss how they depends on the polarization of incident laser light. Later in the Sec.~\ref{subsec:multi-cycle-analysis}, we will ``simulate'' the ESR experiment in which the Li$^+$ ions circulate in the ring, entering and leaving periodically the laser beams. To describe this realistic \textit{multi--cycle} scenario we will apply the rate equations (\ref{eq:rate_equations_general}) in which the induced rates $\Gamma^{{\rm dec, ind}}_{F_3 M_{F_3}, F_i M_{F_i}}$ and $\Gamma^{{\rm exc}}_{F_i M_{F_i}, F_3 M_{F_3}}$ are given by Eq.~(\ref{eq:induced_decay_rate}) for the laser--interaction region and set to \textit{zero} when the ions leave this region.

\subsection{Single--run analysis}
\label{subsec:single_cycle}

After the evaluation of the transition rates $\Gamma^{{\rm dec}}$ and $\Gamma^{{\rm exc}}$, one can integrate the system of coupled equations (\ref{eq:rate_equations_general}) to find the populations of the hyperfine substates $N_{F, M_F}(t)$ as a function of time. Of course, to perform these calculations one has to define first the polarization of both laser beams and the initial population of the Li$^+$ ions. In Fig.~\ref{Fig1_theo}, for example, we display the results obtained for populations that are initially statistically distributed in the ``ground--state'' hyperfine levels:
\begin{subequations}\label{eq:initial_population}
\begin{eqnarray}
    % N_{F_1 M_{F_1}}(0) &\equiv&  N_{\LiTrS: F_1 = 3/2 \, M_{F_1}}(0) = 1/10 \, ,  \\[0.2cm]
    % N_{F_2 M_{F_2}}(0) &\equiv&  N_{\LiTrS: F_2 = 5/2 \, M_{F_2}}(0) = 1/10 \, ,  \\[0.2cm]
    % N_{F_3 M_{F_3}}(0) &\equiv&  N_{\LiTrP{2}: F_2 = 5/2 \, M_{F_3}}(0) = 0 \, ,
    \ketm{\LiTrS   : F_1 = \nicefrac{3}{2}, \, M_{F_1}} & : & \; \; N_{F_1, M_{F_1}}(0) = 1/10 \, ,  \\ %[0.2cm]
    \ketm{\LiTrS   : F_2 = \nicefrac{5}{2}, \, M_{F_2}} & : & \; \; N_{F_2, M_{F_2}}(0) = 1/10 \, ,  \\%[0.2cm]
    \ketm{\LiTrP{2}: F_3 = \nicefrac{5}{2}, \, M_{F_3}} & : & \; \;
    N_{F_3, M_{F_3}}(0) = 0 \, ,
\end{eqnarray}
\end{subequations}
for all $M_{F_i} = - F_i, ..., F_i$. These initial conditions correspond to the scenario where \textit{unpolarized} Li$^+$ ions in the metastable $\LiTrS$ state enter the (laser--ion) interaction region. Moreover, the calculations have been carried out for the horizontally linearly polarized (upper panel of the figure) and circularly polarized $\sigma^{-}-\sigma^{+}$ and $\sigma^{+}-\sigma^{+}$ (middle and lower panels) counter--propagating lasers beams. We have chosen these polarization states in order to reproduce the experimental setup as described in Sec.~\ref{sec:Setup}.

The numerical evaluation of the system (\ref{eq:rate_equations_general}) for the initial conditions (\ref{eq:initial_population}) and for two polarized laser beams allows us to explore the population of all \textit{sixteen} hyperfine substates, involved in the  $2s\,^3\text{S}_1 \, (F_1=\nicefrac{3}{2}) \leftrightarrow 2p\,^3\!{P}_2 \, (F_3=\nicefrac{5}{2}) \leftrightarrow 2s\,^3\!{S}_1 \, (F_2=\nicefrac{5}{2})$ $\Lambda$--transition of interest. For the easier visualization of the results, we do not present in Fig.~~\ref{Fig1_theo} all sixteen $N_{F, M_F}(t)$'s but depict instead the \textit{total} populations $N_{F_i}(t) = \sum_{M_{F_i}} N_{F_i, M_{F_i}}(t)$ of the $\ketm{\LiTrS: F_1 = \nicefrac{3}{2},M_{F_1}}$ (green dash--dotted line line), $\ketm{\LiTrS: F_2 = \nicefrac{5}{2},M_{F_2}}$ (blue dashed line line) and $\ketm{\LiTrP{2}: F_3 = \nicefrac{5}{2},M_{F_3}}$ (black solid line) hyperfine level manifolds. As seen from the figure, the time evolution of the total populations is very sensitive to the polarization state of the lasers. If, for example, both lasers are linearly polarized in the horizontal (storage--ring) plane, all three $N_{F_i}(t)$'s converge to large \textit{non--zero} values at $t \approx 2 \times 10^{-7}$\,s, see upper panel of Fig.~\ref{Fig1_theo}. In this case, approximately 23~\% of Li$^+$ ions from the beam are excited to the $\LiTrP{2} (F_3 = \nicefrac{5}{2})$ level. The spontaneous decay of this level leads to the emission of the $\LiTrP{2} \,(F_3 = \nicefrac{5}{2}) \to \LiTrS\,(F_{1,2} = \nicefrac{3}{2}, \nicefrac{5}{2})$ photons. By making use of density matrix theory \cite{SuJ06}, we found that these photons will be emitted most likely under forward angles with respect to the beam direction and, hence, can be efficiently recorded at the detection region. Since ions with velocity $v = 0.338 \, c$ reach the detection region in about $t_{\rm det} \, \approx \, 10^{-7}$\,s after they enter the laser interaction region, i.e.\ when the population $N_{F_2}$ is almost saturated, our calculations are consistent with the experimental observation of the resonance structure in the spontaneous emission spectrum in Fig.\,\ref{fig:Resonances}.

Qualitatively different behaviour of the populations $N_{F_i}(t)$ is observed if the $\LiTrS \, (F_1=\nicefrac{3}{2}) \leftrightarrow \LiTrP{2} \, (F_3=\nicefrac{5}{2}) \leftrightarrow \LiTrS \, (F_2=\nicefrac{5}{2})$ $\Lambda$--transition is induced by two circularly polarized lasers instead of linearly polarized ones. In order to better illustrate this difference, we solved the system of the rate equations (\ref{eq:rate_equations_general}) for two scenarios in which counter--propagating photons have (i) the opposite $\sigma^{-}-\sigma^{+}$ and (ii) the same $\sigma^{+}-\sigma^{+}$ circular polarization states. The results of the calculations are presented in the middle and lower panels of Fig.~\ref{Fig1_theo}. As seen from these panels, the populations of the hyperfine levels exhibit saturation at later times and at very different values if comparing to the linear polarization case. For example, almost all Li$^{+}$ ions, interacting with oppositely polarized ($\sigma^{-}-\sigma^{+}$) photons, can be found in their $\ketm{\LiTrS: F_2 = \nicefrac{5}{2}}$ level only after $5 \times 10^{-6}$\,s. An even longer time, $t \approx 10^{-4}$\,s, is needed to transfer the entire population to the same $\ketm{\LiTrS: F_2 = \nicefrac{5}{2}}$ level if both lasers are right--handed circularly polarized. The vanishing population in the \LiTrP{2} level can be easily understood from the fact that absorption of circularly polarized $\sigma^{+}-\sigma^{+}$ and $\sigma^{-}-\sigma^{+}$ photons leads to the formation of the \textit{dark} substate $\ketm{\LiTrS: F_2 = \nicefrac{5}{2}, M_{F_2} = - \nicefrac{5}{2}}$, see also Fig.~\ref{Fig_3S1_sublevels}. This substate can \textit{not} be further excited to the $\ketm{\LiTrP{2}: F_3 = \nicefrac{5}{2}}$ level because of missing $M_{F_3}$ substates to fulfill the selection rule $\Delta M_F = - 1$ for $\sigma^+$ red light as indicated in Fig.\,\ref{fig:DarkStates}. As a result, there will be no spontaneous decay $\LiTrP{2} (F_3 = \nicefrac{5}{2}) \to \LiTrS (F_{1,2} = \nicefrac{3}{2}, \nicefrac{5}{2})$ and, hence, no signal at the photomultipliers. Fig.\,\ref{fig:DarkStates} also provides intuitive understanding of the fact that the disappearance of the fluorescence signal is observed later for the $\sigma^{+}-\sigma^{+}$ polarization. In this case photons from the counter--propagating lasers induce $\Delta M_F = +1$ and $\Delta M_F = -1$ transitions and are thus transferring population into opposite directions (lower panel of Fig.\,\ref{fig:DarkStates} and lower pannel of Fig.~\ref{Fig_3S1_sublevels}), whereas their effect is added in the $\sigma^{-}-\sigma^{+}$ case.

The complete transfer of the population to the dark substate $\ketm{\LiTrS: F_2 = \nicefrac{5}{2},M_{F_2} = - \nicefrac{5}{2}}$  happens much later than the time $t \approx 10^{-7}$\,s when the ions reach the detection region, as can be seen from the middle and lower panels of Fig.~\ref{Fig1_theo}. Moreover, our theoretical analysis again showed that during all the (saturation) time the spontaneous decay 
$\LiTrP{2} \,(F_3 = \nicefrac{5}{2}) \to \LiTrS\,(F_{1,2} = \nicefrac{3}{2}, \nicefrac{5}{2})$ leads to the photon emission predominantly in the forward beam direction, i.e.\ under the angles at which the detection region is most sensitive. Based on these observations one would expect that a clear fluorescence signal will be detected for the interaction of Li$^{+}$ ions with circularly polarized lasers. This contradicts, however, the experimental observations, which show no signal for the $\sigma^{-}-\sigma^{+}$ case and a very weak resonance for the $\sigma^{+}-\sigma^{+}$ polarization state in Fig.\,\ref{fig:Resonances}. In order to explain this discrepancy we recall that our calculations have been performed for the initial conditions (\ref{eq:initial_population}), i.e.\ for initially \textit{unpolarized} Li$^{+}$ ions in their ``ground'' levels $\ketm{\LiTrS: \, F_{1,2} = \nicefrac{3}{2}, \nicefrac{5}{2}}$. Such a choice of the initial populations $N_{F, M_F}(t = 0)$ is based on the assumption that \textit{any} non--statistical population of hyperfine levels will be destroyed after ions pass the bending magnets of the ESR storage ring. The observed discrepancy between experimental findings and rate--equation predictions leads us to question this assumption. Namely, the weak or even vanishing fluorescence signal, recorded at the detection region from Li$^{+}$ ions interacting with circularly polarized photons, can be understood if the ions are \textit{already} longitudinally polarized when they enter the laser field. Unless this polarization was produced by the bending magnets, it is caused by the optical pumping during the previous passages of the ion--laser interaction region. 

\subsection{Multi--cycle analysis}
\label{subsec:multi-cycle-analysis}

In order to support our interpretation of \textit{continuous} optical pumping of Li$^{+}$ ions during their circulations in the ESR, we have simulated the \textit{multi--cycle} state-population dynamics. Within one cycle of this simulation, the system of rate equations is used to compute the populations $N_{F, M_{F}}(t)$ in the interval $0 \le t \le t_{\rm ring}$ with $t_{\rm ring} \approx 10^{-6}$\,s being the duration of a single round trip in the ring. Moreover, to account for the fact that ions interact with the laser beams only \textit{part} of their trip we have introduced in Eq.~(\ref{eq:rate_equations_general}) the time--dependent laser--induced (decay and excitation) rates:
\begin{subequations}\label{eq:induced_rate}
\begin{eqnarray}
\Gamma^{\rm exc}(t) &=& 
\begin{cases}
    \Gamma^{\rm exc},& 0 \le t \le t_{\rm laser}\\
    0,               & t_{\rm laser} < t \le t_{\rm ring} 
\end{cases} \, , \\[0.2cm]
\Gamma^{\rm dec, ind}(t) &=& 
\begin{cases}
    \Gamma^{\rm dec, ind},& 0 \le t \le t_{\rm laser}\\
    0,               & t_{\rm laser} < t \le t_{\rm ring} \, ,
\end{cases}
\end{eqnarray}
\end{subequations}
where $t_{\rm laser} \approx 2 \times 10^{-7}$\,s specifies the time during which ions move in the presence of the laser fields. Similar to the $t_{\rm det}$ above, both times $t_{\rm ring}$ and $t_{\rm laser}$ have been estimated for the ESR geometry and for the ion velocity $v = 0.338 c$. 

\begin{figure}[t]
        \includegraphics[width=8.5cm]{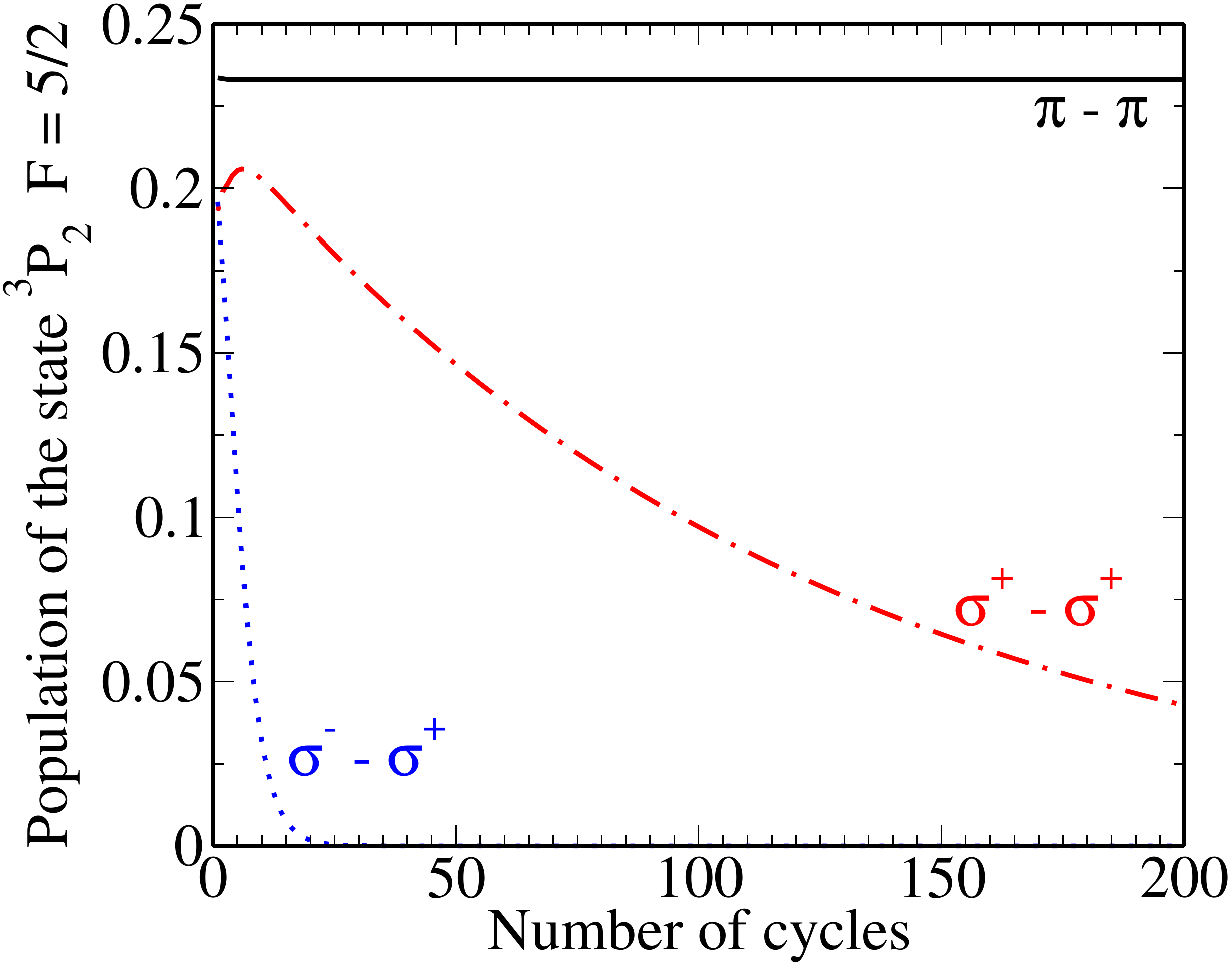}
        \vspace*{0.0cm}
        \caption{The relative population of the excited level $\LiTrP{2}: F_3 = \nicefrac{5}{2}$ of Li$^+$ ions as a function of the number of circulations in the storage ring. $N_{\LiTrP{2}: F = \nicefrac{5}{2}}(t)$ is calculated for the moment when the ions pass the detection region and for the various polarization states of the applied lasers.}
        \label{Fig2_theo}
\end{figure}

The evaluation of the system of rate equations in the time interval $0 \le t \le t_{\rm ring}$ and with the modified induced rates (\ref{eq:induced_rate}) allows us to simulate the state population dynamics during a \textit{single} round trip in the ESR. In order to understand how $N_{F, M_{F}}(t)$'s will change for the realistic experimental scenario of \textit{multiple} circulations of ions in the ring, we (i) employ the populations obtained at the end of a previous cycle, i.e.\ at $t = t_{\rm ring}$, as initial conditions for the next cycle, (ii)  solve the system (\ref{eq:rate_equations_general}) with these conditions, and (iii) repeat the procedure cycle--by--cycle. We have applied this multi--step analysis to investigate how the population of the excited state $\ketm{\LiTrP{2}: F_3 = \nicefrac{5}{2}}$, if observed at the moment when the ions pass the detection region, evolves with the number of cycles. The results of our calculations are displayed in Fig.~\ref{Fig2_theo} for the cases where the Li$^{+}$ ions are pumped by linearly polarized (black solid line) as well as by circularly polarized $\sigma^{-}-\sigma^{+}$ (blue dotted line) and  $\sigma^{+}-\sigma^{+}$ (red dash--dotted line) lasers. We have assumed, moreover, that in the beginning of the very first cycle of the simulation the ions are unpolarized and, hence, their sublevel population is described by Eq.~(\ref{eq:initial_population}).

Figure\,\ref{Fig2_theo} depicts the cycle--by--cycle evolution of the total population $N_{F_3}$ at the time when the ions pass the fluorescence detection region during their multiple turns. It is obvious that the behavior is quite distinct for the different polarizations of the laser beams. If, for example, both lasers are linearly polarized within the ESR plane, the population of the excited level $\ketm{\LiTrP{2}: F_3 = \nicefrac{5}{2}}$ remains virtually constant, $N_{F_3} \approx 0.23$. This is expected from the ``single--run'' results which are presented in the upper panel of Fig.~\ref{Fig1_theo} and indicate that the level populations are saturated to \textit{large} non--zero values in $2 \times 10^{-7}$\,s, i.e.\ approximately at the moment when the ions reach the detection region. Moreover, since the intensity of the $\LiTrP{2}\,(F_3 = \nicefrac{5}{2}) \to \LiTrS (F_{1,2} = \nicefrac{3}{2}, \nicefrac{5}{2})$ spontaneous emission is proportional to $N_{F_3}$, our calculations  support the experimental observation of the pronounced resonance in the fluorescence signal shown in Fig.\,\ref{fig:Resonances}. 

In contrast to the linear--polarization case, the population of the excited level $\ketm{\LiTrP{2}: F_3 = \nicefrac{5}{2}}$ generally decreases with the number of cycles if the ions are pumped by circularly polarized laser beams. One can understand this based on the ``single--run'' analysis from Sec.\,\ref{subsec:single_cycle} as well as from our assumption about the  preservation of the polarization of ions during their round trips. Namely, as seen from Fig.~\ref{Fig1_theo}, the time $t_{\rm laser} \approx 2 \times 10^{-7}$\,s of a single passage of the ions through the laser interaction region is shorter than the  depopulation time of the $\ketm{\LiTrP{2}: F = \nicefrac{5}{2}}$ level, but is sufficient to produce a partial polarization of the Li$^{+}$ ions. If this polarization is preserved---at least to some extent---as the ions continue to move in the ring, the population of the hyperfine substates in the beginning of the next ion--laser interaction period will deviate from Eq.~(\ref{eq:initial_population}). During this next round the lasers will even \textit{further} polarize the ions and the procedure will continue until the formation of the dark substate $\ketm{\LiTrS: F_2 = \nicefrac{5}{2}, M_{F_2} = - \nicefrac{5}{2}}$ is completed and, hence, full depopulation of the $\ketm{\LiTrP{2}: F_3 = \nicefrac{5}{2}}$ level is obtained. 

From our calculations and the discussion above we conclude that the population saturation time in the ``single--run'' scenario is directly related to the number of cycles needed to polarize the ions in the ``multi--cycle'' case. For example, the pumping of Li$^+$ ions by two laser beams with opposite circular polarization, $\sigma^{-}-\sigma^{+}$, results in a relatively fast saturation and, consequently, to the formation of the respective dark substate $\ketm{\LiTrS: F_2 = \nicefrac{5}{2},M_{F_2}= - \nicefrac{5}{2}}$ just after 20 cycles in the ring. In contrast, a very long saturation time, $t \approx 10^{-4}$\,s, and more than 300 round trips are required to polarize Li$^+$ interacting with $\sigma^{+}-\sigma^{+}$ beams. These qualitative predictions, based on the assumption of multi--step optical pumping of Li$^{+}$ ions, allow us to better understand the experimental findings from Fig.\,\ref{fig:Resonances}. Indeed, since for the $\sigma^{-}-\sigma^{+}$ case the ionic population is fully locked in the dark substate $\ketm{\LiTrS: F_2 = \nicefrac{5}{2}, M_{F_2} = - \nicefrac{5}{2}}$ just after a couple of dozend cycles, no signal from the spontaneous decay $\LiTrP{2} (F_3 = \nicefrac{5}{2}) \to \LiTrS (F_{1,2} = \nicefrac{3}{2}, \nicefrac{5}{2})$ can be observed at the detection region. On the other hand, a weak resonance signal recorded for the $\sigma^{+}-\sigma^{+}$ case can be seen as a consequence of a sluggish decrease of the population of the excited state $\ketm{\LiTrP{2}: F_3 = \nicefrac{5}{2}}$ during the circulations of the ions in the ring.    

\begin{figure}[t]
        \includegraphics[width=8.5cm]{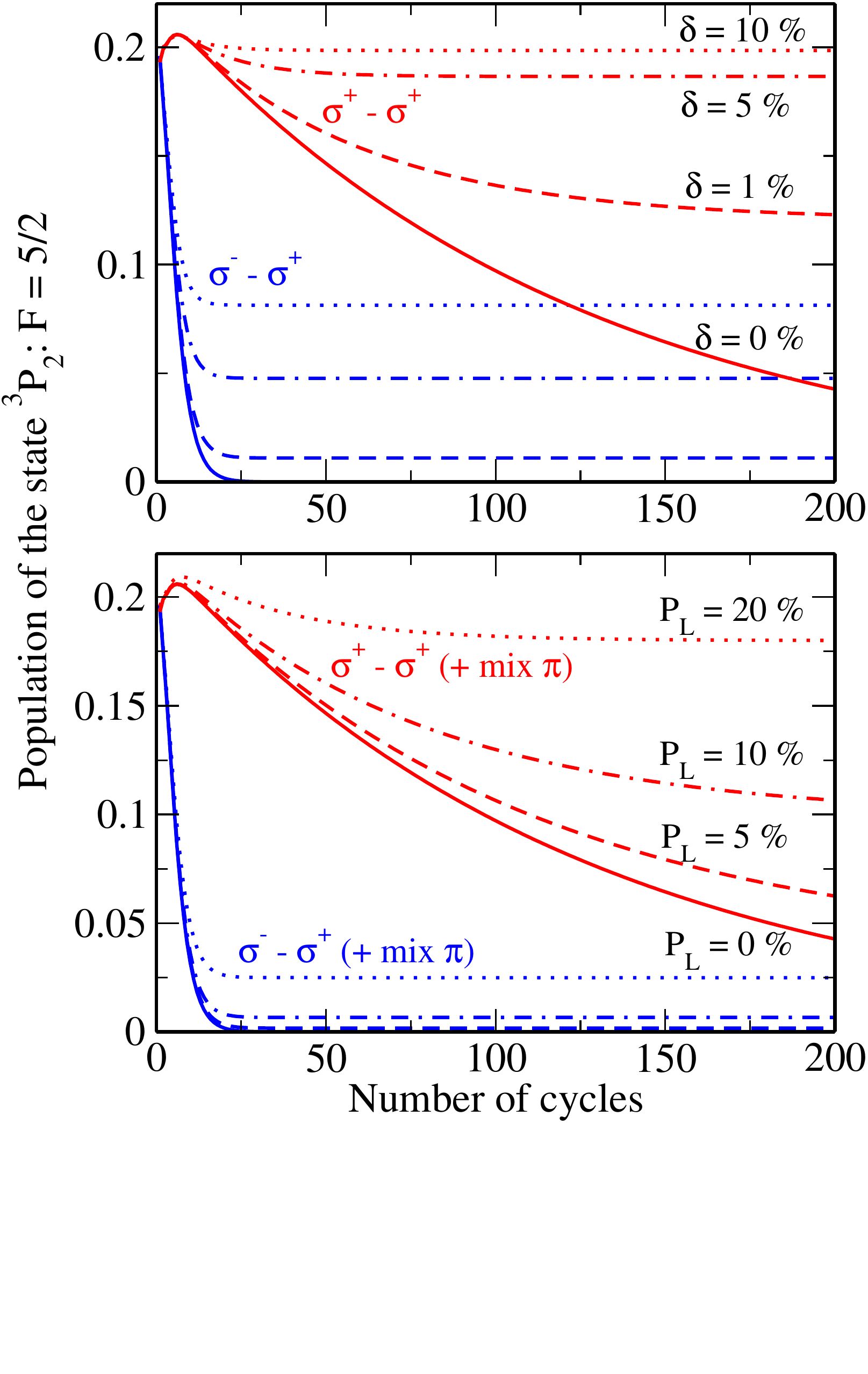}
        \vspace*{-2.5cm}
        \caption{The same as Fig.~\ref{Fig2_theo}, but for the cases where the population of hyperfine substates is partially distorted by the magnets in the storage ring and collisions (upper panel), and the $\sigma$ polarization of the red laser has an admixture of the $\pi$ linearly polarization component (lower panel). In order to estimate the effect of the depolarization in the magnets and collisions, a particular fraction $\delta$ of the population of each substate $\ketm{F M_F}$ was evenly redistributed among ground--level substates after each round trip. The imperfection of the circular polarization is quantified by the Stokes parameter $P_L$, or degree of linear polarization, of the $\pi$ component.}
        \label{Fig3_theo}
\end{figure}

Even though the $\sigma^{+}-\sigma^{+}$ scheme takes longer to transfer the population into the dark substate, a time span of 0.1\,ms, being a fraction of less than 1\% of the dwell time cannot create the remaining signal strength in the spectrum. We shall assume, therefore, that there is a continuous transfer from the dark substate \textit{back} into the multitude of bright states. In order to understand possible reasons for this ``back transfer'' we recall that the theoretical calculations presented in Fig.~\ref{Fig2_theo} are based on two very naive assumptions. Namely, we have presumed that (a) the population of hyperfine substates, as produced during the interaction with lasers and by subsequent spontaneous decay, is neither affected by the magnets of the storage ring nor by collisions and (b) both lasers are ideally ($\sigma^\pm$--) polarized. Both conditions are not fulfilled in the real experiment. Firstly, the polarization of the ions might be modified by collisions with cooler electrons and rest--gas atoms as well as by the magnetic fields of the dipoles and quadrupoles along the circumference of the storage ring. Secondly, the light polarization -- even if it is perfectly circular after the $\nicefrac{\lambda}{4}$ plate -- might obtain a linearly polarized component, e.g., from stress-induced birefringence in the vacuum windows. A detailed analysis of the ion (de--) polarization dynamics due to collisions and interactions with magnetic fields is a rather complicated task which depends, moreover, on the setup of a particular experiment. This analysis is out of scope of the present theoretical study and will be presented in a forthcoming publication. Here we just use a very crude model in which the depolarization of ions in magnets and due to collisions is estimated by redistributing a particular fraction $\delta$ of the population of each $M_F$ state across ground--level substates after each revolution. We found that the system is quite forgiving in the $\sigma^--\sigma^+$ case, where even a depolarization of $\delta = $~10\% per revolution still allows a strong depopulation of the excited level $\LiTrP{2}: F_3 = \nicefrac{5}{2}$ and, hence, a fluorescence signal reduction of about 60~\% after just a dozen cycles, as can be seen in the upper panel of Fig.~\ref{Fig3_theo}. In contrast, in the $\sigma^+-\sigma^+$ case a depolarization of only 1\% leads to a roughly 50~\% signal recovery compared to the calculations neglecting any population redistribution, i.e. for $\delta = 0$~\%. 
As seen from the lower panel of Fig.~\ref{Fig3_theo}, imperfect laser polarization may also lead to the ``back transfer'' from the dark state. Again, it is the $\sigma^--\sigma^+$ case that is more robust: Even with the significant admixture of the $\pi$ component in the red laser beam, quantified by the Stokes parameter $P_L = 20$~\%, the fluorescence signal recovers to only 13\% of its original strength. For $\sigma^+-\sigma^+$, in contrast, already 10\% $\pi$-light is sufficient for recovering 50\% of the signal strength. 
The latter result shows that the small remaining signal in Fig.\,\ref{fig:Resonances} might not necessarily be caused by depolarization effects of the ion beam, but could as well be caused by non-ideal light polarization. More rigorous investigation, experimentally as well as theoretically are required and will be carried out in the future.

\section{\label{sec:Conclusion}Conclusion}
We have shown in our theoretical analysis that optical pumping in the $2s\,^3\!{S}_1\,(F_1=\nicefrac{3}{2}) \, \leftrightarrow \,2p\,^3\!{P}_2\,(F_3=\nicefrac{5}{2}) \leftrightarrow 2s\,^3\!{S}_1 \, (F_2=\nicefrac{5}{2})$ $\Lambda$--transition of Li$^+$ ions at the ESR can only lead to the observed disappearance of the resonance signal if a considerable fraction of the induced polarization survives the revolution along the storage ring. However, with the limited experimental information obtained so far, we cannot conclude that this process has really happened but further investigation of this system is of high interest. Due to the restricted amount of beamtime at the ESR and the excessive number of beamtime applications, we have started to  carry out investigations at CRYRING \cite{Mohr.2019} using Mg$^+$ ions. These can be injected into CRYRING from a local ion source without the necessity of operating and blocking the entire GSI accelerator chain. A $\Lambda$-scheme in $^{25}$Mg$^+$ ions will be addressed with cw lasers but we will also test broad-band optical pumping with pulsed lasers. Those studies serve also for the establishment of laser spectroscopy at CRYRING in general \cite{Lestinsky.2016}. It should be noted that a non-observation of optical pumping at this machine has only limited significance for disproving the explanations of the ESR results due to the completely different lattice and ion dynamics in the two storage rings. Therefore, further experiments at the ESR are also planned. There, a convincing signature could be obtained, \textit{e.g.}, if the resonance disappears again under the conditions of the last experiment but can be reestablished with another laser beam that interacts with the ions on the opposite side of the storage ring and is employed to depopulate the dark--state.
One interesting application of optical pumping at the ESR would be the hydrogen-like ion $^{140}$Pr$^{58+}$ \cite{Frommgen.2019}. Here, laser excitation across the hyperfine splitting in the electronic ground-state, will affect the nuclear lifetime \cite{Litvinov2007}. The direction of neutrino emission can be determined in the Schottky spectrum after the electron capture process \cite{Kienle.2013}. Preferred emission either in forward or in backward direction achieved by optical pumping in the hyperfine transition could then constitute a direct proof of the polarization process.

\section*{Acknowledgement}
We are deeply indebted to the late Dirk Schwalm, who supported and strongly contributed to the SRT experiments for more than two decades.\\
This work was supported by the German Federal Ministry of Education and Research (BMBF, Contract Nos. 06MZ9179I and 05P18RDFAA), and the Helmholtz Association (Contract No. VH-NG-148). G. G. acknowledges support by NSERC (Canada).
\bibliography{OptPump}% Produces the bibliography via BibTeX.

\end{document}